\documentstyle[12pt]{article}

\advance \topmargin by -\headheight
\advance \topmargin by -\headsep
     
\evensidemargin \oddsidemargin
\begin{document}
\newcommand{\sect}[1]{\setcounter{equation}{0}\section{#1}}
\renewcommand{\theequation}{\thesection.\arabic{equation}}

\topmargin -.6in
\def\nonu{\nonumber}
\def\rf#1{(\ref{eq:#1})}
\def\lab#1{\label{eq:#1}} 
\def\br{\begin{eqnarray}}
\def\er{\end{eqnarray}}
\def\be{\begin{equation}}
\def\ee{\end{equation}}
\def\0{\nonumber}
\def\lb{\lbrack}
\def\rb{\rbrack}
\def\({\left(}
\def\){\right)}
\def\v{\vert}
\def\bv{\bigm\vert}
\def\lskip{\vskip\baselineskip\vskip-\parskip\noindent}
\relax
\newcommand{\nit}{\noindent}
\newcommand{\ct}[1]{\cite{#1}}
\newcommand{\bi}[1]{\bibitem{#1}}
\def\a{\alpha}
\def\b{\beta}
\def\ca{{\cal A}}
\def\cm{{\cal M}}
\def\cn{{\cal N}}
\def\cf{{\cal F}}
\def\d{\delta}
\def\D{\Delta}
\def\eps{\epsilon}
\def\g{\gamma}
\def\G{\Gamma}
\def\grad{\nabla}
\def\h{ {1\over 2}  }
\def\hc{\hat{c}}
\def\hd{\hat{d}}
\def\hg{\hat{g}}
\def\hp{ {+{1\over 2}}  }
\def\hm{ {-{1\over 2}}  }
\def\k{\kappa}
\def\l{\lambda}
\def\L{\Lambda}
\def\lg{\langle}
\def\m{\mu}
\def\n{\nu}
\def\o{\over}
\def\om{\omega}
\def\O{\Omega}
\def\p{\phi}
\def\pa{\partial}
\def\pr{\prime}
\def\ra{\rightarrow}
\def\rh{\rho}
\def\rg{\rangle}
\def\s{\sigma}
\def\t{\tau}
\def\th{\theta}
\def\ti{\tilde}
\def\wti{\widetilde}
\def\inte{\int dx }
\def\xb{\bar{x}}
\def\yb{\bar{y}}

\def\tr{\mathop{\rm tr}}
\def\Tr{\mathop{\rm Tr}}
\def\partder#1#2{{\partial #1\over\partial #2}}
\def\ds{{\cal D}_s}
\def\wtwo{{\wti W}_2}
\def\lie{{\cal G}}
\def\alie{{\widehat \lie}}
\def\dlie{{\cal G}^{\ast}}
\def\elie{{\widetilde \lie}}
\def\edlie{{\elie}^{\ast}}
\def\hlie{{\cal H}}
\def\wlie{{\widetilde \lie}}

\def\rlx{\relax\leavevmode}
\def\inbar{\vrule height1.5ex width.4pt depth0pt}
\def\IZ{\rlx\hbox{\sf Z\kern-.4em Z}}
\def\IR{\rlx\hbox{\rm I\kern-.18em R}}
\def\IC{\rlx\hbox{\,$\inbar\kern-.3em{\rm C}$}}
\def\one{\hbox{{1}\kern-.25em\hbox{l}}}

\def\PRL#1#2#3{{\sl Phys. Rev. Lett.} {\bf#1} (#2) #3}
\def\NPB#1#2#3{{\sl Nucl. Phys.} {\bf B#1} (#2) #3}
\def\NPBFS#1#2#3#4{{\sl Nucl. Phys.} {\bf B#2} [FS#1] (#3) #4}
\def\CMP#1#2#3{{\sl Commun. Math. Phys.} {\bf #1} (#2) #3}
\def\PRD#1#2#3{{\sl Phys. Rev.} {\bf D#1} (#2) #3}
\def\PLA#1#2#3{{\sl Phys. Lett.} {\bf #1A} (#2) #3}
\def\PLB#1#2#3{{\sl Phys. Lett.} {\bf #1B} (#2) #3}
\def\JMP#1#2#3{{\sl J. Math. Phys.} {\bf #1} (#2) #3}
\def\PTP#1#2#3{{\sl Prog. Theor. Phys.} {\bf #1} (#2) #3}
\def\SPTP#1#2#3{{\sl Suppl. Prog. Theor. Phys.} {\bf #1} (#2) #3}
\def\AoP#1#2#3{{\sl Ann. of Phys.} {\bf #1} (#2) #3}
\def\PNAS#1#2#3{{\sl Proc. Natl. Acad. Sci. USA} {\bf #1} (#2) #3}
\def\RMP#1#2#3{{\sl Rev. Mod. Phys.} {\bf #1} (#2) #3}
\def\PR#1#2#3{{\sl Phys. Reports} {\bf #1} (#2) #3}
\def\AoM#1#2#3{{\sl Ann. of Math.} {\bf #1} (#2) #3}
\def\UMN#1#2#3{{\sl Usp. Mat. Nauk} {\bf #1} (#2) #3}
\def\FAP#1#2#3{{\sl Funkt. Anal. Prilozheniya} {\bf #1} (#2) #3}
\def\FAaIA#1#2#3{{\sl Functional Analysis and Its Application} {\bf #1} (#2)
#3}
\def\BAMS#1#2#3{{\sl Bull. Am. Math. Soc.} {\bf #1} (#2) #3}
\def\TAMS#1#2#3{{\sl Trans. Am. Math. Soc.} {\bf #1} (#2) #3}
\def\InvM#1#2#3{{\sl Invent. Math.} {\bf #1} (#2) #3}
\def\LMP#1#2#3{{\sl Letters in Math. Phys.} {\bf #1} (#2) #3}
\def\IJMPA#1#2#3{{\sl Int. J. Mod. Phys.} {\bf A#1} (#2) #3}
\def\AdM#1#2#3{{\sl Advances in Math.} {\bf #1} (#2) #3}
\def\RMaP#1#2#3{{\sl Reports on Math. Phys.} {\bf #1} (#2) #3}
\def\IJM#1#2#3{{\sl Ill. J. Math.} {\bf #1} (#2) #3}
\def\APP#1#2#3{{\sl Acta Phys. Polon.} {\bf #1} (#2) #3}
\def\TMP#1#2#3{{\sl Theor. Mat. Phys.} {\bf #1} (#2) #3}
\def\JPA#1#2#3{{\sl J. Physics} {\bf A#1} (#2) #3}
\def\JSM#1#2#3{{\sl J. Soviet Math.} {\bf #1} (#2) #3}
\def\MPLA#1#2#3{{\sl Mod. Phys. Lett.} {\bf A#1} (#2) #3}
\def\JETP#1#2#3{{\sl Sov. Phys. JETP} {\bf #1} (#2) #3}
\def\JETPL#1#2#3{{\sl  Sov. Phys. JETP Lett.} {\bf #1} (#2) #3}
\def\PHSA#1#2#3{{\sl Physica} {\bf A#1} (#2) #3}
\def\PHSD#1#2#3{{\sl Physica} {\bf D#1} (#2) #3}
\begin{titlepage}
\vspace*{-2 cm}
\noindent
\begin{flushright}

\end{flushright}

\vskip 1 cm
\begin{center}
{\Large\bf Torsionless T-selfdual Affine NA  Toda Models  \footnote{talk given at the VI International Wigner symposiun, Istambul,
Turkey, 1999}} \vglue 1  true cm
{ J.F. Gomes}$^{\dagger}$, E. P. Gueuvoghlanian$^{\dagger}$
 { G.M. Sotkov}\footnote{On leave of absence from the Institute for Nuclear
 Research and Nuclear Energy, Bulgarian Academy of Sciences, 1784, Sofia,
 Bulgaria}$^{\dagger}$ and { A.H. Zimerman}$^{\dagger}$\\

\vspace{1 cm}

$^{\dagger}${\footnotesize Instituto de F\'\i sica Te\'orica - IFT/UNESP\\
Rua Pamplona 145\\
01405-900, S\~ao Paulo - SP, Brazil}\\
 jfg@ift.unesp.br, gueuvogh@ift.unesp.br, sotkov@ift.unesp.br, zimerman@ift.unesp.br\\

\vspace{1 cm}

\end{center}

\normalsize
\vskip 0.2cm

\begin{center}
{\large {\bf ABSTRACT}}\\
\end{center}
\noindent

A general construction of affine  Non Abelian Toda models  in terms of gauged
two loop WZNW model is discussed.  In particular we find the
 Lie algebraic condition defining  a subclass of {\it T-selfdual
torsionless NA Toda models} and  their zero curvature representation.

\noindent

\vglue 1 true cm

\end{titlepage}

\section{Introduction}
Two dimensional integrable models represent  an important laboratory
for testing new ideas and developing new methods for constructing exact
solutions as well as for the nonperturbative quantization of 4-D non-abelian
gauge theories, gravity and  string theory.  Among the numerous
technics for constructing 2-D integrable models and their solutions
\cite{fadeev}, \cite{lez-sav}, the two loop $\hat \lie$ -WZNW and gauged $\hat
G / \hat H$-WZNW models \cite{Aratyn} have the advantage of providing a simple
and universal method for derivation of the zero curvature representation as
well as a consistent path integral formalism for their description.  The power
of such method was demonstrated in constructing (multi) soliton solution of the
abelian affine Toda models\cite{Aratyn}
 and certain nonsingular nonabelian (NA) affine Toda
models \cite{luis}.  

The present paper is devoted to the sistematic construction of the simplest
class of singular torsionless affine NA Toda models characterized 
 by the fact that
the space of physical fields $g_0^{f}$ lies in the coset 
$ \lie_0 /{\lie_0^0} = {{SL(2)}\o {U(1)}} \otimes 
U(1)^{rank \lie -1}$.  Our  {\it main result} is that such models exists only for
the following  three affine  Kac-Moody
algebras, $B_n^{(1)}, A_{2n}^{(2)} $ and $ D_{n+1}^{(2)}$ under certain 
specific  restrictions on the choice of the subgroup $\lie_0^0$ (i.e.
equivalently the choice of gradation $Q$ and constant grade $\pm 1$ elements
$\eps_{\pm}$).  It turns out that these models are T-selfdual (i.e. the 
axial and the
vector gauging of the $U(1)$  factor in the coset ${{SL(2)}\o {U(1)}} \otimes 
U(1)^{rank \lie -1}$ leads to identical actions ).  They appear to be natural
generalization of the Lund-Regge (``complex Sine-Gordon'') model \cite{lund} 
and exactly
reproduce the  family of models proposed by Fateev
\cite{Fat}.  Our construction provide  a simple proof of their classical
integrability.

It is important to mention 
that relaxing the  structure of the coset 
 $ \lie_0 /{\lie_0^0} ={{SL(2) \otimes 
U(1)^{rank \lie -1}}\o {U(1)}} $, i.e. gauging specific 
combinations of the Cartan
subalgebra of $ \lie_0$ we find two new families of integrable models,
 {\it axionic}
(for axial gauging) and {\it torsionless} ( for vector gauging) for all affine
(twisted and untwisted ) Kac-Moody algebras which are T-dual (but not self
dual)\cite{emilio}.   

An important  motivation for the construction of the above singular NA Toda 
models is the fact that their soliton solutions (for imaginary coupling
)carries both electric and magnetic (topological ) charges and have properties
quite similar to the 4-D dyons of the Yang-Mills-Higgs model \cite{emilio}.


The generic NA Toda models  are classified
 according to a $\lie_0 \subset \lie$ embedding  induced
by the grading operator $Q$ decomposing an finite or infinite Lie algebra 
$\lie = \oplus _{i} \lie _i $ where $
[Q,\lie_{i}]=i\lie_{i}$ and $ [\lie_{i},\lie_{j}]\subset \lie_{i+j}$.  A group
element $g$ can then be written in terms of the Gauss decomposition as 
\be
g= NBM
\label{1}
\ee
where $N=\exp \lie_< \in H_{-}$, $B=\exp \lie_{0} $ and
$M=\exp \lie_> \in H_{+}$.  The physical fields $B$ 
lie in the zero grade subgroup $\lie_0$ 
and the
models we seek correspond to the coset $H_- \backslash G/H_+ $.

For consistency with the hamiltonian reduction formalism, the phase space of
the G-invariant WZNW model is  reduced by specifying the constant
generators $\eps_{\pm}$ of grade $\pm 1$.  In order to derive 
 an action for $B \in \lie_0 $,  invariant under 
\begin{eqnarray}
g\longrightarrow g^{\prime}=\alpha_{-}g\alpha_{+},
\label{2}
\end{eqnarray}
where $\a_{\pm}(z, \bar z) \in H_{\pm}$. 
 we have to introduce a set of  auxiliary
gauge fields $A \in \lie _{<} $ and $\bar A \in \lie _{>}$ transforming as 
\begin{eqnarray}
A\longrightarrow A^{\prime}=\alpha_{-}A\alpha_{-}^{-1}
+\alpha_{-}\partial \alpha_{-}^{-1},
\quad \quad 
\bar{A}\longrightarrow \bar{A}^{\prime}=\alpha_{+}^{-1}\bar{A}\alpha_{+}
+\bar{\partial}\alpha_{+}^{-1}\alpha_{+}.
\label{3}
\end{eqnarray}
The result is 
\begin{eqnarray}
S_{G/H}(g,A,\bar{A})&=&S_{WZNW}(g)
\nonumber
\\
&-&\frac{k}{2\pi}\int dz^2 Tr\( A(\bar{\partial}gg^{-1}-\epsilon_{+})
+\bar{A}(g^{-1}\partial g-\epsilon_{-})+Ag\bar{A}g^{-1}\) .
\nonumber
\end{eqnarray}
Since the action $S_{G/H}$ is $H$-invariant,
 we may choose $\alpha_{-}=N_{-}^{-1}$
and $\alpha_{+}=M_{+}^{-1}$.  From the orthogonality  of the graded 
subpaces, i.e. $Tr \lie _ij =0, i+j \neq 0$, we find 
\begin{eqnarray}
S_{G/H}(g,A,\bar{A})&=&S_{G/H}(B,A^{\prime},\bar{A}^{\prime})
\nonumber
\\
&=&S_{WZNW}(B)-\frac{k}{2\pi}
\int dz^2 Tr[A^{\prime}\epsilon_{+}+\bar{A}^{\prime}\epsilon_{-}
+A^{\prime}B\bar{A}^{\prime}B^{-1}],
\label{14}
\end{eqnarray}
where 
\begin{eqnarray}
S_{WZNW}=- \frac{k}{4\pi }\int d^2zTr(g^{-1}\partial gg^{-1}\bar{\partial }g)
-\frac{k}{24\pi }\int_{D}\epsilon_{ijk}
Tr(g^{-1}\partial_{i}gg^{-1}\partial_{j}gg^{-1}\partial_{k}g),
\label{3a}
\end{eqnarray}
and the topological term denotes a surface integral  over a ball $D$
identified as  space-time.

Action (\ref{14}) describe the non singular Toda models among which we find the
Conformal and the Affine abelian Toda models where 
$Q=\sum_{i=1}^{r}\frac{2\lambda_{i}\cdot H}{\alpha_{i}^{2}}, \quad 
 \epsilon_{\pm}=\sum_{i=1}^{r}c_{\pm i}E_{\pm \alpha_{i}}$ and 
$Q= h d + \sum_{i=1}^{r}\frac{2\lambda_{i}\cdot H}{\alpha_{i}^{2}}, \quad 
\epsilon_{\pm}=\sum_{i=1}^{r}c_{\pm i}E_{\pm \alpha_{i}}^{(0)} + 
E_{\mp \psi }^{(\pm 1)}$
respectively, where $\psi $ denote the highest root, $\lambda_i$,  the 
fundamental weights, $h$ the coxeter number of $\lie $
  and $H_i$  are the Cartan subalgebra generators 
in the Cartan
Weyl basis, $ Tr (H_iH_j) = \d_{ij}$.

 More interesting cases 
arises in
connection with non abelian embeddings $\lie_0 \subset \lie $. 
 In particular, if we 
supress one
of the fundamental weights from $Q$, the zero grade subspace acquires 
a nonabelian 
structure
$sl(2)\otimes u(1)^{rank \lie -1}$.  Let us consider for instance 
$Q= h^{\pr} d + \sum_{i\neq a}^{r}\frac{2\lambda_{i}\cdot H}{\alpha_{i}^{2}}$, 
where $h^{\pr} =0$ or
$h^{\pr} \neq 0$ corresponding to
 the Conformal and Affine  nonabelian (NA) Toda 
respectively.
The absence of $\lambda_a$ in $Q$
 prevents the contribution of the simple root step operator
$E_{\a_a}^{(0)}$ in constructing $\eps_+$. It in fact, allows for 
reducing the phase space 
even further.  This fact can be understood by enforcing the 
nonlocal constraint $J_{Y \cdot H} = \bar J_{Y \cdot H} = 0$
where $Y$ is such that $[Y\cdot H , \eps_{\pm}] = 0$ and 
$J=g^{-1}\partial g$ and $\bar{J}=-\bar{\partial}gg^{-1}$.  Those generators 
of $\lie_0$ commuting with
$\eps_{\pm}$ define a subalgebra $\lie_0^0 \subset \lie_0 $.
 Such subsidiary constraint is incorporated into the action by
  requiring symmetry under \cite{annals} 
\begin{eqnarray}
g\longrightarrow g^{\prime}=\alpha_{0}g\alpha_{0}'
\label{5}
\end{eqnarray} 
where we shall consider   $\a^{\pr}_{0}
 =\alpha_{0}(z, \bar z) \in \lie_0^0 $, i.e., {\it axial symmetry} (the 
 {\it vector }
 gauging is obtained by choosing $\a^{\pr}_{0}
 ={\alpha_{0}}^{-1}(z, \bar z) \in \lie_0^0 $). 
Auxiliary gauge fields $A_0 = a_0 Y\cdot H$ and $\bar A_0= 
\bar a_0 Y\cdot H\in  \lie_{0}^{0}$  are 
introduced by splitting
$A = A_- + A_0$ and  $\bar A = \bar A_+ + \bar A_0$ transforming as
\begin{eqnarray}
A\longrightarrow A^{\prime}=\alpha_{0}A\alpha_{0}^{-1}
+\alpha_{0}\partial \alpha_{0}^{-1}, \quad \quad 
\bar{A}\longrightarrow \bar{A}^{\prime}=\alpha_{0}^{-1}\bar{A}\alpha_{0}
+\bar{\partial}\alpha_{0}^{-1}\alpha_{0},
\nonu 
\end{eqnarray}
\begin{eqnarray}
A_{0}\longrightarrow A_{0}^{\prime}=A_{0}+\alpha_{0}^{-1}\partial \alpha_{0},
\quad \quad 
\bar{A}_{0}\longrightarrow \bar{A}_{0}^{\prime}=\bar{A}_{0}
+\bar{\partial}\alpha_{0}\alpha_{0}^{-1}.
\nonu 
\end{eqnarray}

The invariant action, under transformations generated by
$H_{(-,0)}$ (in left) and $ H_{(+,0)}$ (in right) is given by
\begin{eqnarray}
S_{G/H}(g,A,\bar{A})&=&S_{WZNW}(g)-\frac{k}{2\pi}\int d^2z 
Tr \( A(\bar{\partial}gg^{-1}-\epsilon_{+}) \right.
\nonu
\\
&+& \left. \bar{A}(g^{-1}\partial g-\epsilon_{-})+
Ag\bar{A}g^{-1}+A_{0}\bar{A}_{0} \) .
\label{45}
\end{eqnarray}
Notice that the physical fields
 $g_0^f$ lie in the coset $\lie_0 /{\lie_0^0} = {{sl(2)\otimes
u(1)^{rank \lie -1}}/u(1)}$ of dimension $rank
\lie +1$ and are classified according to the 
gradation $Q$.  It therefore follows that $S_{G/H}(g,A,\bar{A})=
 S_{G/H}(g_0^f,A^{\pr},\bar{A}^{\pr})$.

In \cite{ime} a detailed study of the gauged WZNW construction 
for  finite dimensional Lie algebras leading to Conformal NA Toda
models was presented.  The study of its symmetries was given in refs. \cite{plb} and in
\cite{annals}.  Here we generalize the construction of ref. \cite{ime} to
infinite dimensional Lie algebras leading to NA Affine 
Toda models characterized 
by the broken
conformal symmetry and by the  presence of
solitons. 
 
Consider the Kac-Moody
algebra ${\widehat \lie }$
\br
[T_m^a ,T_n^b]  =  f^{abc} T^c_{m+n} + {\hat c}m \d_{m+n} \d^{ab} \nonu 
\er
\br
[{\hat d} , T^a_n] = nT^a_n ;\quad [{\hat c}, T^a_n] = [{\hat c},{\hat d} ] = 0
\label{km}
\er

The NA Toda models we shall be constructing are associated to 
gradations of the type\footnote{ We are considering 
$\lie ^{\pr}$ to be
semisimple and therefore $a$ to be one
of the end points of the Dynkin diagram of $\lie $, otherwise the model
decomposes in two abelian Toda models coupled by $\psi $ and $\chi$.}.
$Q_a(h^{\pr}) = h^{\pr} d + 
\sum_{i\neq a}^{r}\frac{2\lambda_{i}\cdot H}{\alpha_{i}^{2}}$, 
where $h^{\pr}$
is chosen such that the zero grade subalgebra $\hat \lie_0$ defined by 
 $Q_a(h^{\pr})$, coincide  with the    zero grade 
subalgebra $\lie_0$ defined by  $Q_a(h^{\pr}=0)$ 
 (apart from two  comuting 
 generators $\hat {c}$ and $\hat {d}$).  Since they commute with $\lie_0$,
  the kinetic part
 decouples such that the conformal and the affine singular NA-Toda  models 
 differ only by the
 potential term. 
Due to  the specific 
graded structure of the algebra, the following  trace properties 
hold
\br
&&{\rm Tr}A \bar{\partial}g_{0}^f{(g_{0}^f)}^{-1}=
{\rm Tr}A_{0}\bar{\partial}g_{0}^f{(g_{0}^f)}^{-1},
\quad \quad
{\rm Tr}\bar{A}{(g_{0}^f)}^{-1}\partial g_{0}^f=
{\rm Tr}\bar{A}_{0}{(g_{0}^f)}^{-1}\partial g_{0}^f
\nonumber
\\
&&{\rm Tr}Ag_{0}^f\bar{A}^{\prime}{(g_{0}^f)}^{-1}=
{\rm Tr}A_{0}g_{0}^f\bar{A}_0{(g_{0}^f)}^{-1}+
{\rm Tr}A_{-}g_{0}^f\bar{A}_{+}{(g_{0}^f)}^{-1}.
\nonumber
\er
Henceforth the action decomposes into three parts, i. e.,
\begin{eqnarray}
S_{G/H}=S_{WZNW}(g_{0}^{f})+F_{0}+F_{\pm},
\label{48}
\end{eqnarray}
where
\begin{equation}
F_{0}=-\frac{k}{2\pi}\int d^2z Tr\( A_{0}\bar{\partial}g_{0}^{f}
({g_{0}^{f}})^{-1}
+\bar{A}_{0}({g_{0}^{f}})^{-1}\partial g_{0}^{f}
+A_{0}g_{0}^{f}\bar{A}_{0}({g_{0}^{f}})^{-1}+A_{0}\bar{A}_{0} \),
\nonu 
\end{equation}
\begin{eqnarray}
F_{\pm}=-\frac{k}{2\pi}\int dz^2 Tr\( A_{-}\hat {\epsilon_{+}}
+\bar{A}_{+}\hat {\epsilon_{-}}
+A_{-}g_{0}^{f}\bar{A}_{+}({g_{0}^{f}})^{-1} \),
\nonu 
\end{eqnarray}
and the functional integral now factorizes into 
\begin{eqnarray}
Z=\int DA_{0}D\bar{A}_{0}\exp (-F_{0})\int DA_{-}D\bar{A}_{+}\exp (-F_{\pm}).
\label{51}
\end{eqnarray}
The integration over the auxiliary gauge fields $A$ and $\bar A$ require 
explicit
parametrization of $B$. 
\begin{eqnarray}
B=\exp (\tilde {\chi} E_{-\alpha_{a}})
 \exp (   R {{Y^j}} {{h_j}}+\Phi (H)+ \nu \hat {c} +
  \eta \hat {d})\exp (\tilde {\psi} E_{\alpha_{a}})
 \label{63}
 \end{eqnarray}
where $ \Phi (H) =\sum_{i=1}^{r-1}\varphi_{i}{{X}^j}_i {{h_j}}$
,  $\sum_{j=1}^r{{Y^j}}  {{X}^j}_{i} =0$ and $ h_i = {{2 \a_i\cdot H} \o {\a^2}}, 
 i=1, \cdots ,r-1$.
After gauging away the nonlocal field $R$, the factor group element becomes    
\be
g_0^f=\exp (\chi E_{-\alpha_{a}})
 \exp (   \Phi (H)+ \nu \hat {c} + \eta \hat {d})\exp (\psi E_{\alpha_{a}})
 \label{63a}
\ee
where $\chi = \tilde {\chi}e^{{1\o 2}Y\cdot \a_a}, \quad 
\psi = \tilde {\psi}e^{{1\o 2}Y\cdot \a_a}$
We therefore get for the zero grade component
\be
F_0 ={-{k\o {2\pi}}}\int \( a_0 \bar a_0 2Y^2\Delta +  2({{\a_a \cdot Y}\o
{\a_a^2}})(\bar a_0\psi \pa \chi + a_0 \chi \bar \pa \psi )e^{\Phi (\a_a)}\)d^2x
\label{del}
\ee
where $\Delta = 1 + {{(Y \cdot \a_a )^2}\o {2Y^2}}\psi \chi e^{\Phi (\a_a )}$ 
and  
\be
F_{\pm} = {-{k\o {2\pi}}}\int \(Tr
 (A_- - g_0^f\hat {\eps_-} (g_0^f)^{-1})g_0^f(\bar A_+ 
 - (g_0^f)^{-1}\hat {\eps_+} g_0^f) 
 (g_0^f)^{-1}\)d^2x
\label{fmm}
\ee
The effective action is obtained by integrating over the auxiliary fields $A_0, \bar A_0, A_-$
and $\bar A_+$, 
\be
Z_0 = \int DA_{0}D\bar{A}_{0}\exp (F_{0}) \sim e^{-S_0}
\ee
where $S_0 = -{k \o {\pi}}({{Y \cdot \a_a }\o {\a_a^2}})\int {{\psi \chi \bar
 \pa \psi \pa \chi }\o
{Y^2 \Delta }}e^{2\Phi(\a_a)}$.   Also, 
\be
Z_{\pm} = \int DA_{-}D\bar{A}_{+}\exp (F_{\pm}) \sim  e^{-{k \o {2\pi}}\int
 Tr \hat {\eps_+} g_0^f\hat {\eps_-} (g_0^f)^{-1}}
 \ee
 The total action (\ref{45}) is therefore given as
 \be
 S= -{k \o {4\pi}}\int \( Tr (\pa \Phi(H)\bar \pa \Phi(H)) + 
 {{2\bar \pa \psi \pa \chi }\o \Delta
 }e^{\Phi(\a_a)} + \pa \eta \bar \pa \nu +
 \pa \nu \bar \pa \eta 
    - 2 Tr (\hat {\eps_+} g_0^f\hat {\eps_-} (g_0^f)^{-1}) \)
 \label{action}
 \ee

Note that the second term in (\ref{action}) contains both symmetric and
antisymmetric parts:
\begin{eqnarray}
\frac{e^{\Phi(\a_a)} } {\Delta}\bar{\partial}\psi \partial \chi
=\frac{ e^{\Phi(\a_a)} }
{\Delta}(g^{\mu \nu}\partial_{\mu}\psi\partial_{\nu}\chi
+\epsilon_{\mu \nu}\partial_{\mu}\psi \partial_{\nu}\chi ),
\end{eqnarray}
where $g_{\mu \nu}$ is the 2-D metric of signature $ g_{\mu
\nu}= diag (1,-1)$, $\pa = \pa_0 + \pa_1\;\; \bar \pa = \pa_0 - \pa_1$.
 For $n=1$ ($\lie \equiv A_{1}$, $\Phi (\a_1)$ is zero) the
antisymmetric term is a total derivative:
\begin{eqnarray}
\epsilon_{\mu \nu}\frac{\partial_{\mu}\psi \partial_{\nu}\chi}{1+\psi \chi}
=\frac{1}{2}\epsilon_{\mu \nu}\partial_{\mu}
\left( \ln \left\{ 1+\psi \chi \right\}
\partial_{\nu}\ln{\frac{\chi}{\psi}}\right),
\end{eqnarray}
and it can be neglected.  This $A_{1}$-NA-Toda model (in the conformal case), 
is known to describe the
2-D black hole solution for (2-D) string theory \cite{Witten1}.
The
$G_n $-NA conformal Toda model can be used in the
description of specific (n+1)-dimensional black string theories 
\cite{gervais-saveliev},
 with  n-1-flat and
2-non flat directions ($g^{\mu
\nu}G_{ab}(X)\partial_{\mu}X^{a}\partial_{\nu}X^{b}$, $X^{a}=(\psi ,\chi
,\varphi_{i})$), containing axions ($\epsilon_{\mu
\nu}B_{ab}(X)\partial_{\mu}X^{a}\partial_{\nu}X^{b}$) and tachions
($\exp \left\{ -k_{ij}\varphi_{j}\right\} $), as well.

It is clear that the presence of the $ e^{\Phi(\a_a)}$ in (\ref{action})
is responsible for the
 antisymmetric  tensor
generating the axionic terms. 
On the other hand, notice that $\Phi(\a_a)$ depend upon the subsidiary nonlocal
constraint $J_{Y \cdot H} = \bar J_{Y \cdot H} = 0$ and hence upon the choice of
the vector {\bf{Y}}.  It is defined to be orthononal to all roots contained in
$\eps_{\pm}$.

 In ref. \cite{ime}, the most general 
constant grade one element $\eps_+$
was analysed and the precise condition for the absence of 
axions, i.e. $\Phi(\a_a)=0 $, was established determining a
subclass of torsionless conformal 
NA singular Toda models (no torsion theorem).
For finite dimensional Lie algebras, it was shown in ref. \cite{ime} that the
absence of axions can only occur for $G_n = B_n$, $a=n$ and and $\eps_{\pm} = 
\sum_{i=1}^{n-2}c_{\pm i}E_{\pm \alpha_{i}}
+d_{\pm}E_{\pm (\alpha_{n}+\alpha_{n-1})}$.  In such case, $\lie_0^0$ is
generated by $Y\cdot H = (\frac{2\lambda_{n}}{\alpha_{n}^{2}}
-\frac{2\lambda_{n-1}}{\alpha_{n-1}^{2}})\cdot H$ and 
$ \Phi (H) =\sum_{i=1}^{n-2}\varphi_{i}h_i
+  \varphi_{-}(\a_{n-1}+\a_n ) \cdot H $.
Due to the root structure of $B_n$, we verify that $ \Phi (\a_n) = \a_n \cdot
(\a_{n-1} + \a_n)\varphi_{-}=0$


In extending the no torsion theorem to infinite affine Lie algebras, 
   $h^{\pr}$ is chosen by defining the gradation
$Q_a(h^{\pr})$ such that preserves
$\lie_0$.  
 We consider $\hat \eps_+ = \eps_+ + E_{-\psi}^{(1)}$ where 
$\psi $ is the highest root of $\lie$.  Since conformal and the affine models differ only
by the potential term, the solution for the no torsion condition is also
satisfied for infinite dimensional algebras, whose Dynkin diagram possess a
$B_n$-``tail like''.
An obvious solution  is the untwisted $B_n^{(1)}$ model. Two other  solutions
were found within the twisted affine Kac-Moody algebras $A_{2n}^{(2)}$ and
$D_{n+1}^{(2)}$ as we shall describe in detail.

\section{The $B_n^{(1)}$ Torsionless NA Toda model}

Let $Q = 2(n-1) d + 
\sum_{i=1}^{n-1}\frac{2\lambda_{i}\cdot H}{\alpha_{i}^{2}}$ decomposing
$B_n^{(1)}$ into graded subspaces. 
 In particular $\lie_0 = SL(2) \otimes
U(1)^{n-1}\otimes U(1)_{\hat c}\otimes  U(1)_{\hat d}$ generated by
$\{E_{\pm \a_n}^{(0)}, h_1, \cdots , h_n,{\hat c}, {\hat d} \}$. 
Following
the no torsion theorem of ref. \cite{ime}, we have to choose  
 $\hat {\eps_{\pm}} = \sum_{i=1}^{n-2} c_{\pm i}E_{\pm \a_i}^{(0)} +
 c_{\pm (n-1)}E_{\pm (\a_{n-1}+\a_n)}^{(0)} + c_{\pm n}E_{\mp \psi }^{(\pm 1)}$, 
 where $\psi = \a_1 + 2(\a_2 + \cdots + \a_{n-1}
 + \a_n)$ is the highest root of $B_n$ and $\lie_0^0$ is
generated by $Y\cdot H = (\frac{2\lambda_{n}}{\alpha_{n}^{2}}
-\frac{2\lambda_{n-1}}{\alpha_{n-1}^{2}})\cdot H$ 
 such that $[{\bf{Y \cdot H}}, \hat {\eps_{\pm}}] = 0$.  The coset $\lie_0
 /\lie_0^0$ is then parametrized according to (\ref{63a}) with $\Phi(H) =
 \sum_{i=1}^{n-1} {\cal H}_i \varphi_i + \eta \hat h + \nu \hat c $
where ${\cal H}_i = (\a_n + \cdots \a_i )\cdot H$ so that 
$Tr ({\cal H}_i {\cal H}_j) = \d_{ij}, i,j=1, \cdots , n-1$ and 
the total
effective action becomes
\be
S=-\frac{k}{4\pi}\int d^{2}x
\( \frac{1}{4}\sum_{i=1}^{n-1}g^{\mu \nu}\pa_{\mu }\varphi_{i}\pa
_{\nu}\varphi_{i}+
g^{\mu \nu }\frac{\pa _{\mu}\psi \pa _{\nu}\chi}{1+\psi \chi} +
\frac{1}{2}g^{\mu \nu} \pa_{\mu} \nu \pa _{\nu} \eta - 2V \)
\label{affaction}
\ee
where the ``affine potential'' $(n>2)$ is 
\be
V=\sum_{i=1}^{n-2}|c_i|^2 e^{\varphi_{i}-\varphi_{i+1}}+
 |c_{n-1}|^2 2(1+2\psi \chi )e^{-\varphi_{n-1}}
+|c_{n}|^2e^{\varphi_{1}+\varphi_{2}-\eta}
\label{affV}
\ee
The action (\ref{affaction}) is  invariant under conformal
tranformation 
\br
z\rightarrow f(z),\quad \bar z\rightarrow g(\bar z),
\quad \psi \rightarrow \psi ,\quad \chi \rightarrow \chi ,
\nonumber 
\er
\br
\varphi_{s}\rightarrow \varphi_{s}+s\ln f^{\prime}g^{\prime};
\quad s=1,2,...,n-1;
\quad \eta \rightarrow \eta +2(n-1)\ln f^{\prime}g^{\prime}
\er
We should point out that 
 the $\eta$ field plays a crucial role in establishing the conformal
invariance of the theory.  Integrable deformation of such class of theories
can  than, be sistematically obtained by setting $\eta =0$.

For the case $n=2$ we choose,  $ \hat {\eps_{\pm}} = E_{\a_1 +\a_2 }^{(0)} 
+E_{-\a_1 -\a_2 }^{(1)}$, $\Phi(\a_{n-1}) = \varphi $, i.e. 
${\widehat \lie } =
{\widehat SO}(5)$, is also special in the sense that its
 complexified theory,
i.e.
\br
\psi \longrightarrow i\psi; \quad \chi \longrightarrow i\psi^* ; \quad
\varphi \longrightarrow  i\varphi  \nonu
\er
leads to the real action  
\be
S=-\frac{k}{4\pi}\int d^{2}x
\( {{g^{\mu \nu}\pa _{\mu}\psi \pa _{\nu}\psi^* }\o {(1-\psi \psi^* )}}+
{1\o 4}g^{\mu \nu } \pa _{\mu }\varphi  \pa _{\nu} \varphi 
+ 8(1-2\psi \psi^* )cos \varphi \) 
\label{so5action}
\ee
\section{ The twisted  NA Toda Models}

The twisted affine Kac-Moody algebras are constructed from a finite
dimensional algebra possessing a nontrivial symmetry of their Dynkin
diagrams (folding).  Such symmetry can be extended to the algebra by 
  an outer automorphism $\sigma $ \cite{cornwell}, 
   as 
\be
\s ( E_{\a}) = \eta_{\a} E_{\s (\a)}
\label{aut}
\ee
where $\eta_\a = \pm 1$.  For the simple roots, $\eta_{\a_i} = 1$.  The
signs can be consistently assign to all generators since  
nonsimple roots can be written as sum of  
 two roots other roots.

 The no torsion theorem  require a $B_n$-``tail like'' structure which is
 fulfilled only by the $A_{2n}^{(2)}$ and $D_{n+1}^{(2)}$ (see appendix N of ref.
 \cite{cornwell}).  In both cases
 the automorphism is of order 2 (i.e. $ \s ^2 =1 $). 
 
 Let us denote by $\a$ the roots of the untwisted algebra $\lie$.  
 For the $A_{2n}^{(2)}$ case, the automorphism is defined by 
 \be
 \s ({\a_1}) = {\a_{2n}}, \;\; \s ({\a_2}) = {\a_{2n-1}}
\;\; \cdots ,\s ({\a_{n-1}}) = {\a_{n}}
\label{fol-a2n}
\ee
whilst for  the $D_{n+1}^{(2)}$, the automorphism acts only in the
``fish tail'' of the Dynkin diagram of $D_{n+1}$, i.e. 
\be 
\s (E_{\a_1}) = E_{\a_{1}}, \;\;\cdots ,\s (E_{\a_{n-1}}) = E_{\a_{n-1}},
\;\;\s (E_{\a_{n}}) = E_{\a_{n+1}}
\label{fol-dn}
\ee

The automorphism $\s$  decomposes the algebra $\lie = \lie _{even}
\bigcup \lie _{odd}$. The twisted affine algebra is
constructed from $\lie $ assigning an affine index $m \in Z$
 to the generators in $\lie _{even}$ while $m \in Z + {1\o 2}$ to those in 
 $\lie _{odd}$ (see appendix N of \cite{cornwell}).
 
The simple root step operators for $A_{2n}^{(2)}$     are  
\be
E_{\b_i} = E^{(0)}_{\a_i} + E^{(0)}_{\a_{2n-i+1}}, \;\; i= 1, \cdots ,n 
 \;\;\;\;
E_{\b_0} =  E^{({1\o 2})}_{-\a_1 - \cdots - \a_{2n}} 
\label{simplea2n}
\ee
corresponding to the 
  simple and highest roots
 \be
\b_i = {1\o 2} (\a_i + \a_{2n-i+1}) \;\; i= 1, \cdots ,n,  \;\; \;\; 
 \psi = \a_1 + \cdots + \a_{2n} = 2(\b_1 + \cdots \b_n)
 \label{simprootsa2n}
 \ee
  respectively.
  
  For $D_{n+1}^{(2)}$,  simple root step operators are
\br
E_{\b_i} = E^{(0)}_{\a_i}, \;\; i= 1, \cdots ,n-1, \;\;\;
E_{\b_n} = E^{(0)}_{\a_n} + E^{(0)}_{\a_{n+1}} \nonu 
\er
\br
E_{\b_0} =  E^{({1\o 2})}_{-\a_1 - \cdots - \a_{n-1} - \a_{n+1}}-
E^{({1\o 2})}_{-\a_1 - \cdots - \a_{n-1} - \a_{n}}
\label{simpledn}
\er
corresponding to the 
  simple and highest roots
 \br
\b_i =  \a_i \;\; i= 1, \cdots ,n-1,  \;\; \;\;
\b_n = {1\o 2} (\a_n + \a_{n+1}), 
\nonumber 
\er
\be
 \psi = \a_1 + \cdots \a_{n-1} + {1\o 2}(\a_{n} + \a_{n+1})=  
 \b_1 + \cdots \b_n
 \label{simplrootsdn}
 \ee
 where have denoted by $\b$ the roots of the twisted (folded) algebra.
 
 The torsionless affine NA Toda models are defined by 
 \be
 Q = 2(2n-1)\hat d + \sum_{i \neq n, n+1}^{2n}{{2\lambda_i \cdot H}\o {\a_i^2}},
 \label{qa2n}
 \ee
 and 
 \be
 Q = (2n-2)\hat d + \sum_{i=1}^{n}{{2\lambda_i \cdot H}\o {\a_i^2}}
 \label{qdn}
 \ee
 for  $A_{2n}^{(2)}$ and $D_{n+1}^{(2)}$ respectively, where $\lambda_i $ are
 the fundamental weights of the untwisted algebra $\lie $, i.e. ${{2\lambda_i
 \cdot \a_j }\o {\a_j^2}} = \d_{ij}$.
 
 Both models are specified by the constant grade $\pm 1$ operators $\eps_{\pm}$
 \be
 \hat {\eps_{\pm}} = \sum_{i=1}^{n-2} c_{\pm i}E_{\pm \b_i} + c_{\pm (n-1)}E_{\pm
 (\b_{n-1}+\b_n)} + c_{\pm n} E_{\mp \b_0}
 \label{eps}
 \ee
 where $\b_i$ are the simple roots of the twisted affine algebra specified in
 (\ref{simprootsa2n}) and  in (\ref{simplrootsdn}).
 
 According to the grading generators (\ref{qa2n}) and (\ref{qdn}), the zero
 grade subalgebra is in both cases $\lie_0 = 
 SL(2)\otimes U(1)^{n-1}\otimes U(1)_{\hat c}\otimes  U(1)_{\hat d}$ generated
 by $\{ E^{(0)}_{\pm \b_n}, h_1, \cdots , h_n, \hat c, \hat d \}$. 
  Hence the zero grade subgroup is
 parametrized as in (\ref{63}) where we have taken $\eta =0$, responsible for 
 breaking
  the conformal invariance.  The   factor group is given in
  (\ref{63a}), where
 $\lie_0^0 $ is generated by 
 $Y\cdot H = (\frac{2\mu_{n}}{\b_{n}^{2}}
-\frac{2\mu_{n-1}}{\b_{n-1}^{2}})\cdot H$ and    $\mu_i$ are the fundamental
 weights  of the twisted algebra i.e. ${{2\mu_i \cdot \b_j} \o {\b_j^2}}=
 \d_{ij}$
 In order to decouple the $\varphi_i, \;\; i=1, \cdots , n-1$ 
 we chose an orthonormal basis for the Cartan subalgebra, i.e.
  $\Phi (H) = {\cal H}_i \varphi_i + \eta \hat h + \nu \hat c $
where
\be 
{\cal H}_i  = ( \a_i + \cdots \a_{2n-i+1})\cdot H, \;\;
   Y\cdot H = {\cal H}_n,  \;\; 
 Tr ({\cal H}_i{\cal H}_j) = 2\d_{ij}, \; i,j = 1, \cdots n
\label{cartana2n}
\ee
 and 
\be 
 {\cal H}_i = (\a_{n-i+1}  + \cdots + \a_{n+1})\cdot H, \;\;
  Y\cdot H = {\cal H}_n,   \;\;
 Tr ({\cal H}_i{\cal H}_j) = \d_{ij}, \; i,j = 1, \cdots n
\label{cartandn}
\ee
for   $A_{2n}^{(2)}$ and $D_{n+1}^{(2)}$ respectively.

The Lagrangean density is obtained from (\ref{action}) leading to 
\be
{\cal L}_{A_{2n}^{(2)}} = {{\pa \chi \bar \pa \psi }\o {1+ {1\o 2}\psi \chi }}
+ {1\o 2} \sum_{i=1}^{n-1}\pa \varphi_i \bar \pa  \varphi_i - V_{A_{2n}^{(2)}}
\label{acta2n}
\ee
and
\be 
{\cal L}_{D_{n+1}^{(2)}} = 2{{\pa \chi \bar \pa \psi }\o {1+ \psi \chi }}
+ {1\o 2} \sum_{i=1}^{n-1}\pa \varphi_i \bar \pa  \varphi_i - V_{D_{n+1}^{(2)}}
\label{actdn}
\ee
where 
\be
V_{A_{2n}^{(2)}} = \sum_{i=1}^{n-2}|c_i|^2 e^{-\varphi_i + \varphi_{i+1}} +
{1\o 2}|c_n|^2 e^{2\varphi_1} + |c_{n-1}|^2e^{-\varphi_{n-1}}(1+ \psi \chi )
\label{pota2n}
\ee
and 
\be
V_{D_{n+1}^{(2)}} = \sum_{i=1}^{n-2}|c_i|^2 e^{-\varphi_i + \varphi_{i+1}} +
{1\o 2}|c_n|^2 e^{\varphi_1} + |c_{n-1}|^2e^{-\varphi_{n-1}}(1+ 2\psi \chi )
\label{potdn}
\ee

The models described by  (\ref{affaction}), (\ref{acta2n}) and (\ref{actdn}) 
coincide with those proposed by Fateev in \cite{Fat}.

\section{Zero Curvature}
The equations of motion for the NA Toda models are known to be of the form
\cite{leznov-saveliev}
\be 
\bar \pa (B^{-1} \pa B) + [\hat {\eps_-}, B^{-1} \hat {\eps_+} B] =0, 
\quad \pa (\bar \pa B B^{-1} ) - [\hat {\eps_+}, B\hat {\eps_-} B^{-1}] =0
\label{eqmotion}
\ee
 The subsidiary constraint $J_{Y \cdot H} = Tr(B^{-1} \pa B Y\cdot H)=
  \bar J_{Y \cdot H} = Tr(\bar \pa B B^{-1}Y \cdot H )=        0$ can be
 consistenly imposed  since $[Y\cdot H, \hat {\eps_{\pm}}]=0$ as can be 
 obtained from
 (\ref{eqmotion}) by taking the trace with $Y.H$.  Solving those equations 
 for the nonlocal
 field $R$ yields, 
 \be
 \pa R = ({{Y\cdot \a_n}\o {Y^2}}) {{\psi \pa \chi }\o \Delta }e^{\Phi(\a_n)},
 \;\; \;\;\; 
\bar \pa R = ({{Y\cdot \a_n}\o {Y^2}}) {{\chi \pa \psi }\o \Delta }e^{\Phi(\a_n)}
\label{bh}
\ee
The equations of motion for the fields $\psi, \chi $ and $\varphi_i, i=1,
\cdots , n-1$  obtained from (\ref{eqmotion}) imposing the constraints
(\ref{bh}) coincide precisely with the Euler-Lagrange equations derived from
(\ref{acta2n}) and (\ref{actdn}).  Alternatively, (\ref{eqmotion}) admits a
zero curvature representation $\pa \bar A - \bar \pa A + [A, \bar A] =0$ where
\be
A= \hat {\eps_-} + B^{-1} \pa B,\quad  
\bar A= -B^{-1}\hat {\eps_+}  B
\label{zcc}
\ee
Whenever the constraints (\ref{bh}) are incorporated into $A$ and $\bar A$ in
(\ref{zcc}), equations (\ref{eqmotion}) yields the zero curvature
representation of the NA singular Toda models.  Such argument is valid for all
NA Toda models, in particular for the torsionless class of models discussed in
the previous two sections.

Using the explicit parametrization of $B$ given in (\ref{63}),  the
corresponding $\hat {\eps_{\pm}}$ specified in (\ref{eps}), 
 (\ref{simprootsa2n}) and (\ref{simplrootsdn}) together with  
(\ref{bh}) where $Y$ is given in 
 (\ref{cartana2n}) and (\ref{cartandn}), we find, in
a systematic manner, the following form for $A$ and $\bar A$
\br
A_{A_{2n}^{(2)}} &= & \sum_{i=1}^{n-2} c_i (E^{(0)}_{-\a_i} +
E^{(0)}_{-\a_{2n-i+1}}) +  
 c_{n-1} (E^{(0)}_{-\a_n - \a_{n-1}} + E^{(0)}_{-\a_{n+1}-\a_{n+2}})\nonu \\ 
  &+&
 c_n  E^{(-{1\o 2})}_{ \a_1 + \cdots + \a_{2n}}
 + 
 \pa \psi e^{-{1\o 2}R}(E^{(0)}_{\a_n} + E^{(0)}_{\a_{n+1}}) 
+\sum_{i=1}^{n-1} \pa \varphi_i {\cal H}_i \nonu \\
&+& 
{{\pa \chi }\o \Delta }e^{{1\o 2}R}(E^{(0)}_{-\a_n} + E^{(0)}_{-\a_{n+1}})
\label{zcca2n}
\er
and 
\br
-\bar A_{A_{2n}^{(2)}} &=& \sum_{i=1}^{n-2} c_i e^{-\varphi_i + \varphi_{i+1}}
(E^{(0)}_{\a_i} +
E^{(0)}_{\a_{2n-i+1}}) + c_n e^{2\varphi_1}
 E^{({1\o 2})}_{ -\a_1 - \cdots - \a_{2n}}\nonu \\ 
&+& c_{n-1}e^{-\varphi_{n-1}}(E^{(0)}_{\a_n + \a_{n-1}} +
 E^{(0)}_{\a_{n+1}+\a_{n+2}})\nonu \\
&+&c_{n-1}\psi e^{-{1\o 2}R -\varphi_{n-1}}(E^{(0)}_{\a_{n+1}+\a_n + \a_{n-1}} -
 E^{(0)}_{\a_n+ \a_{n+1}+\a_{n+2}})\nonu \\ 
&+& c_{n-1}\chi e^{{1\o 2}R -\varphi_{n-1}}(E^{(0)}_{ \a_{n-1}} -
 E^{(0)}_{\a_{n+2}}) 
+ c_{n-1}\psi \chi e^{-\varphi_{n-1}}(E^{(0)}_{\a_n + \a_{n-1}} -
 E^{(0)}_{ \a_{n+1}+\a_{n+2}}) \nonu \\ 
&+& {1\o 2}c_{n-1}\psi ^2 \chi e^{-\varphi_{n-1}-{1\o 2}R}
(E^{(0)}_{\a_{n+1}+\a_n + \a_{n-1}} -
 E^{(0)}_{\a_n+ \a_{n+1}+\a_{n+2}}) 
\label{zcca2nbar}
\er
\br
A_{D_{n+1}^{(2)}} &= & \sum_{i=1}^{n-2} c_i E^{(0)}_{-\a_i} 
 +  
 c_{n-1} (E^{(0)}_{-\a_n - \a_{n-1}} + E^{(0)}_{-\a_{n-1}-\a_{n+1}})\nonu \\ 
  &+&
 c_n  (E^{(-{1\o 2})}_{ (\a_1 + \cdots + \a_{n-1}+ \a_{n+1})} -
 E^{(-{1\o 2})}_{ (\a_1 + \cdots + \a_{n}+ \a_{n+1})})
 + 
 \pa \psi e^{-{1\o 2}R}(E^{(0)}_{\a_n} + E^{(0)}_{\a_{n+1}}) \nonu \\
&+&\sum_{i=1}^{n-1} \pa \varphi_i {\cal H}_i 
+ 
{{\pa \chi }\o \Delta }e^{{1\o 2}R}(E^{(0)}_{-\a_n} + 
E^{(0)}_{-\a_{n+1}})
\label{zccdn}
\er
and 
\br
-\bar A_{D_{n+1}^{(2)}} &=& \sum_{i=1}^{n-2} c_i e^{-\varphi_i + \varphi_{i+1}}
E^{(0)}_{\a_i}   
+ c_{n-1}e^{-\varphi_{n-1}}(E^{(0)}_{\a_n + \a_{n-1}} +
 E^{(0)}_{\a_{n+1}+\a_{n-1}})\nonu \\
&+& 2c_{n-1}\psi e^{-{1\o 2}R -\varphi_{n-1}}E^{(0)}_{\a_{n+1}+\a_n
+\a_{n-1} }
+ 2c_{n-1}\chi e^{{1\o 2}R -\varphi_{n-1}}E^{(0)}_{ \a_{n-1}}\nonu \\  
&+& 2c_{n-1}\psi \chi e^{-\varphi_{n-1}}(E^{(0)}_{\a_{n+1} + \a_{n-1}} 
 + E^{(0)}_{\a_{n-1} + \a_{n}}) 
+ c_{n-1}\psi ^2 \chi e^{-{1\o 2}R -\varphi_{n-1}}
E^{(0)}_{\a_{n+1}+\a_n + \a_{n-1}} \nonu \\
 &+& c_{n+1}e^{\varphi_1}
(E^{({1\o 2})}_{ -(\a_1 + \cdots + \a_{n-1}+ \a_{n+1})} -
 E^{({1\o 2})}_{ -(\a_1 + \cdots + \a_{n}+ \a_{n+1})}) 
\label{zccdnbar}
\er

For the untwisted affine $B^{(1)}_n$ model of the previous section the zero
curvature representation is obtained from
\br
A_{B_{n}^{(1)}} &= & \sum_{i=1}^{n-2} c_i E^{(0)}_{-\a_i}
+  
 c_{n-1} E^{(0)}_{-\a_n - \a_{n-1}} 
+c_n  E^{(-{1})}_{ \a_1 +2(\a_2 + \cdots +  \a_{n})}\nonu \\  
&+& \pa \psi e^{-{1\o 2}R}E_{\a_n}^{(0)} 
+\sum_{i=1}^{n-1} \pa \varphi_i {\cal H}_i 
+{{\pa \chi }\o \Delta }e^{{1\o 2}R } E_{-\a_n}^{(0)}
\label{zccbn}
\er
\br
-\bar A_{B_{n}^{(1)}} &=& \sum_{i=1}^{n-2} c_i e^{-\varphi_i + \varphi_{i+1}}
E^{(0)}_{\a_i}    + c_n e^{\varphi_1 + \varphi_2 }
E^{(1)}_{ -(\a_1 +2(\a_2 + \cdots +  \a_{n}))} 
+ 2 \chi e^{\varphi_{n-1}+ {1\o 2}R}E_{-\a_{n-1}}^{(0)} \nonu \\
&+& c_{n-1}(1+2 \psi \chi )e^{-\varphi_{n-1}}E^{(0)}_{\a_{n-1} + \a_n} 
-2c_{n-1}e^{-\varphi_{n-1} -{1\o 2}R} \psi (1+ \psi \chi )E^{(0)}_{\a_{n-1} 
+ 2\a_n}
\label{zccbnbar}
\er

The zero curvature representation of such  subclass of torsionless NA Toda
models shows that they are in fact classically integrable field theories. 
The  construction of the previous sections provides a  sistematic affine Lie
algebraic structure underlying those models.

\section{Conclusions}

We have constructed a class of affine NA Toda models from the gauged two-loop 
WZNW models
in which left and right symmetries are incorporated by a suitable choice of
grading operator $Q$.  Such framework is specified by grade $\pm 1$ constant
 generators
$\eps_{\pm}$ and the pair $(Q,\eps_{\pm} )$ determines the model in terms of a
zero grade subgroup $\lie_0$.  We have shown that for non abelian $\lie_0$, it
is possible to reduce even further the phase space by constraining to
zero the currents 
commuting with $\eps_{\pm}$ , ($J \in \lie_0^0)$) to  the fields lying in the
coset $\lie_0/\lie_0^0$ only.  Moreover, we have found  a Lie algebraic condition 
 which defines a class of {\it T-selfdual
torsionless models}, for the case $\lie_0^0 = U(1)$.  The action  
for those models were sistematicaly 
constructed and shown to coincide with the models
proposed by Fateev \cite{Fat}, describing the {\it strong coupling limit} of specific 
2-d models representing
sine-Gordon interacting with Toda-like models. Their {\it weak coupling limit}
appears to be  the Thirring model coupled to certain affine Toda theories 
\cite{Fat}.

Following the same line of arguments of the previous sections, one can construct
more general models, say,  $ \lie_0 /{\lie_0^0} = 
{{SL(2)\otimes U(1)^{n-1}}\o {U(1)^{s}}}$, 
$ \lie_0 /{\lie_0^0} = 
{{SL(2)\otimes SL(2)\otimes U(1)^{n-2}} \o {U(1)}}$, 
$ \lie_0 /{\lie_0^0} = 
{{SL(3)\otimes U(1)^{n-2}}\o {U(1)}}$, etc.  Those models represent more general NA
affine Toda models obtained by considering specific gradations $Q_{a,b,\cdots }
= h_{a,b, \cdots } d + 
\sum_{i\neq a,b \cdots }^{n}\frac{2\lambda_{i}\cdot H}{\alpha_{i}^{2}}$. 
However the important problem of the classification of all integrable models
obtained as gauged two loop $G$-WZNW models remains open.

{\bf Acknowledgments}  We are grateful to FAPESP and CNPq for 
financial support.

\end{document}